\documentclass{PoS}
\usepackage{unites2e}
\usepackage{siunitx}
\usepackage{amsmath}
\usepackage{subfigure}
\title{A detailed study of Gamma-ray emission from Cassiopeia A using VERITAS}

\ShortTitle{Gamma-ray emission from Cassiopeia A}

\author{\speaker{Sajan Kumar$^{1}$ for VERITAS collaboration}\thanks{for the VERITAS collaboration.}\\
        $^1$University of Delaware, 19716, USA\\
        E-mail: \email{sajan@udel.edu}}

\abstract{
Supernova remnants (SNRs) have long been considered the leading candidate sites for the acceleration of cosmic rays within the Galaxy through the process of diffusive shock acceleration. The connection between SNRs and cosmic rays is supported by the detection of high energy (HE; 100 MeV to 100 GeV) and very high energy (VHE; $100\U{GeV}$ to 100 TeV) gamma rays from young and middle-aged SNRs. However, the interpretation of the gamma-ray observations is not unique. This is due to the fact that gamma rays can be produced by electrons through non-thermal Bremsstrahlung and inverse Compton scattering, and by protons through proton-proton collisions and subsequent neutral pion decay. To disentangle and quantify the contributions of electrons and protons to the gamma-ray flux, it is necessary to measure precisely the spectra and morphology of SNRs over a broad range of gamma-ray energies. Cassiopeia A (Cas A) is one such young SNR (~350 years) which is bright in radio and X-rays. It has been detected as a bright point source in HE gamma rays by Fermi-LAT and in VHE gamma rays by HEGRA, MAGIC and VERITAS. Cas A has been observed with VERITAS for more than 60 hours - three times the published exposure. The observations were taken between 2007 and 2013 over a wide range of zenith angles. In particular, half of the total data were taken under large zenith angles to boost the effective area above few TeV. Here, we will present the detailed spectral and morphological results from the complete dataset. 

}

\FullConference{The 34th International Cosmic Ray Conference,\\
		30 July- 6 August, 2015\\
		The Hague, The Netherlands}

\begin{document}

\section{Introduction}
For decades, young supernova remnants(SNRs) have been thought to be responsible for the production of the bulk of Galactic cosmic rays through the mechanism of diffusive shock acceleration. The support for this conviction is backed by a number of observational facts such as the detection of non-thermal X-rays and high energy/very high energy gamma rays from these SNRs\cite{Aharonian:2013gp}. In the class of SNRs, Cassiopeia A (Cas A) is one of the youngest (age $\sim$ 350 yrs \cite{Ashworth:1980cas}) object, located at a distance of ($\sim$ $3.4 \U{kpc}$ \cite{Reed:1995cas}). Observations of the scattered light echo from the supernova explosion indicate that Cas A was a type IIb supernova from a star of initial mass 15-25 $M_{\odot}$ and had lost most of its hydrogen envelop prior to exploding \cite{Krause:2008cas}. Due to its proximity to Earth and its well established progenitor and brightness, it has been observed extensively across the electromagnetic spectrum, from radio to gamma rays. Viewed in radio wavelengths \cite{Bell:1975cas, Baars:1977cas}, it is one of the brightest sources, with the main fraction of the emission coming from a bright radio ring (radius $\sim$ 100") thought to be associated with the reverse shock. This radio emission is generated through the synchrotron process by electrons that are trapped and accelerated in the magnetic field. Optical emission is dominated mainly by dense ejecta heated by reverse-shock and several fast moving knots \cite{Fesen:1988cas}. Broadening of CO line emission towards Cas A has been observed by the Spitzer space telescope, which suggests an interaction between the shock front of the SNR and molecular clouds \cite{Hines:2004cas, Kilpatrick:2014cas}. In X-rays ($\sim$ 0.1 keV to 80 keV), thermal and non-thermal emission is seen with \textit{XMM-Newton}, \textit{Chandra} and \textit{NuSTAR}. Thermal X-ray emission arises from the shell of reverse-shocked ejecta, rich in emission lines from highly ionized atoms and overlapping with the bright radio shell \cite{Fabian:1980cas, Laming:2003cas}. Besides this, non-thermal emission is also seen at both forward and reverse shock \cite{Gotthelf:2001cas, Uchiyama:2008cas} , although the emission at the reverse shock can also be interpreted as the emission from the forward shock projected towards the inner ring \cite{DeLaney:2004cas}. Recently, observations from NuSTAR resolve the remnant above 15 KeV and find that emission is dominated not only by forward and reverse shocks but also by knots located in the interior of remnant \cite{Grefenstette:2014cas}. This non-thermal X-ray emission can be interpreted as synchrotron radiation from electrons accelerated  to a few tens of TeV within shocks. \par
 Since electrons are being accelerated to relativistic energies at both forward and reverse shocks, it has been argued that gamma-rays (from MeV to TeV range) should be expected from this source through inverse Compton scattering. However, it is difficult to differentiate the contribution of the forward and reverse shocks to gamma-ray emission because of the limited angular resolution of space-based and ground-based gamma-ray observatories. Alternatively, if protons are accelerated at the shock fronts, they can interact with ambient target material (gas), also producing gamma-ray emission through neutral pion decay. Recently, direct evidence for the acceleration of protons at SNR shocks has been confirmed from two SNRs: IC443 and W44 \cite{Ackermann:2013gp}. \par
  In the HE gamma-ray  energy range (MeV to GeV),  the \textit{Fermi} Gamma ray space Telescope detected a signal from Cas A for the first time in 2010 \cite{Fermi:2010cas}, but did not provide very strong constraints on the emission models. A more recent paper by Yuan et al. \cite{Yuan:2013cas} shows that a hadronic model is preferred over leptonic model in the GeV energy range and this point is supported by Saha et al. \cite{Saha:2014casa} also.  In TeV regime, HEGRA  first discovered VHE gamma-ray emission from Cas A \cite{HEGRA:2001cas}, with later confirmations of this detection supplied by MAGIC and VERITAS observatories \cite{MAGIC:2003cas, VERITAS:2010cas}. The spectral index ($\gamma = 2.4 - 2.6$) and the fluxes(~$3\%$ of the Crab Nebula flux) published by these three groups match well with each other within errors. Due to the limited angular resolution of these instruments, Cas A is detected as a point source of gamma rays in the TeV energy range.\par
  In this work, we will present the results from observations of Cas A with VERITAS taken between 2007 and 2013,  which amount to more than 60 hours. This exposure is almost three times that already published by VERITAS, and significantly reduces the statistical error on the flux and spectral index, providing further constraints which help to distinguish between hadronic and leptonic models of gamma-ray emission. Here we present a brief summary of the VERITAS instrument, followed by the procedures for data selection and analysis, and discuss the results and conclusions.

\section{VERITAS Telescope}

VERITAS (the Very Energetic Radiation Imaging Telescope Array System) is a ground-based gamma-ray observatory which consists of an array of four telescopes, located in southern arizona at (\ang{30;40;30.21}N, \ang{110;57;07.77}W). Each telescope has a 12-m diameter optical reflector, providing a total reflecting area of  $\sim$ 110$\si{\square\meter}$.  A camera consisting of 499 photomultiplier tubes (PMTs) is placed in the focal plane of each telescope, giving a field of view of $3.5\si{\degree}$. VERITAS employs the technique of stereoscopic reconstruction of gamma-ray initiated air showers. From 2007 to 2012, the array underwent two major upgrades. The first occurred during the summer of 2009, when the original prototype telescope was moved to a new location to make the whole array layout more symmetric and improve the baseline distances. In the second upgrade, in 2011-2012, FPGA-based L2 trigger system and higher efficiency PMTs (super-bialkali photocathod) were installed \cite{Zitzer:2013gp, Kieda:2013gp} . Along with the inclusion of new data analysis techniques, these upgrades have resulted in improved sensitivity of the array and a lower energy threshold. Currently, a source with a flux level of 1$\%$ of the Crab Nebula can be detected in 25h. The angular resolution for gamma-rays at $1\U{TeV}$ energy is $\sim 0.08\si{\degree}$ and the sensitive energy detection range spans from $85\U{GeV}$ to $>30\U{TeV}$.

\section{Data selection and Analysis}
The observations of Cas A are summarised in Table 1. All observations were performed using four telescopes under very dark and clear sky conditions. After quality selection cuts the total exposure is $\sim$65 hours. The first step in the analysis procedure is image calibration and cleaning \cite{Cogan:2006gt}. This step is necessary to select only those pixels which contain the Cherenkov light and remove all other pixels containing night sky background. Calculation of Hillas parameters (length, width and size of image) is the second step \cite{Hillas:1985gp}. From these parameters, a gamma ray initiated shower can be distinguished from a background shower initiated by cosmic rays. After the image parametrization, the arrival direction of each event is reconstructed using the stereoscopic technique, wherein the major axes of  the shower images intersect at a point in the camera plane, providing a geometrical method to measure the arrival direction of an event.

\begin{table*}
 \caption{Details of observations of Cas A.}
 \vspace{5mm}
 \centering
 \begin{tabular} {c c c c c c c c }
 \hline\hline
 Data Set & Date & N$_{tels}$ & $\theta_Z$ range & $Average \theta_{Z}$ & Wobble &Live Time&Mean trigger rate \\
 & & &(deg)&(deg)&(deg)&(Hours) & (Hz)\\
 \hline
 \textrm{I} & 09/07 - 11/07 & 4 &27-40&34&0.5&18& 250 \\
 \textrm{II} & 12/11 - 12/11 & 4 &33-43&38&0.5&2& 350 \\
 \textrm{III}& 09/12 - 12/13 & 4 &24-39&30&0.5&19&400 \\
 \textrm{IV} & 09/12 - 12/13 & 4 &40-64&56&0.5&25&300 \\
 
  \hline
 \end{tabular}
 \label {table:Crab}
 \end{table*}

\section{Results}

\begin{figure}[ht]
\begin{center}
\begin{tabular}{cc}
\includegraphics [width=70mm, height=55mm]{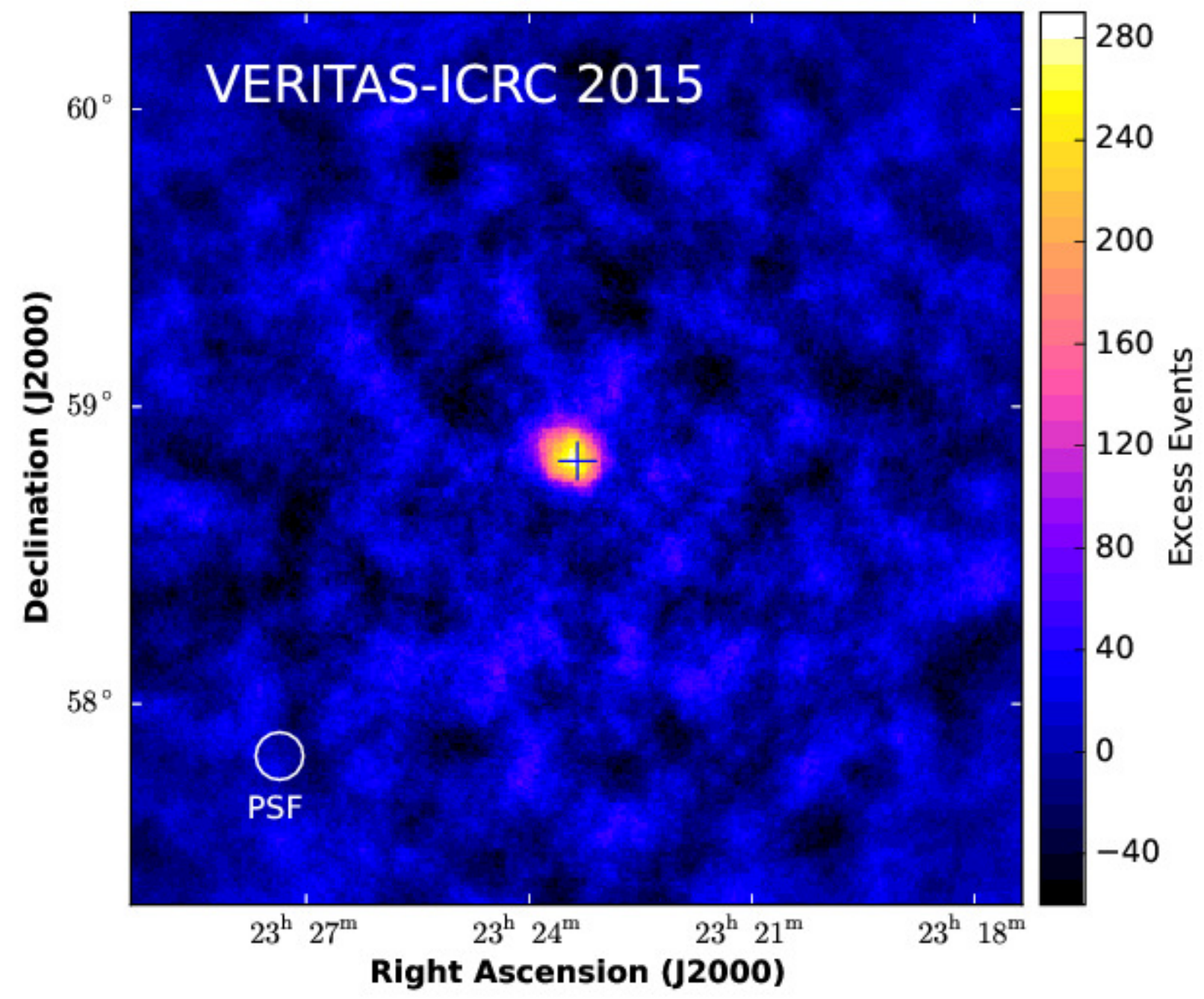}
\quad
\includegraphics [width=55mm, height=55mm]{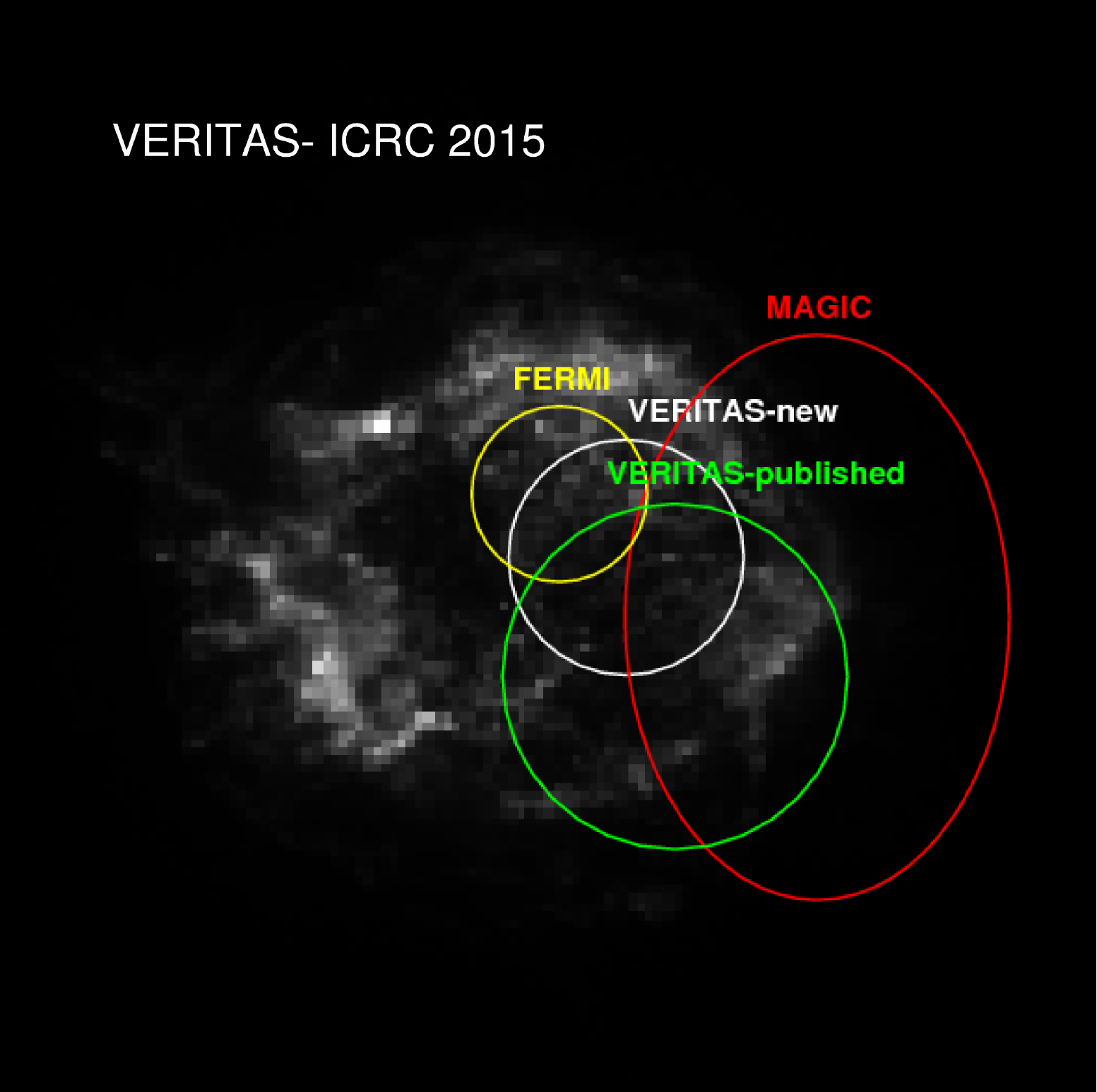}
\end{tabular}
\end{center}

\caption{\protect \small (left) The skymap of excess counts from the region of Cas A, smoothed with a circular window of radius 0.09\si{\degree}. This map was produced using 18 hours of VERITAS observations from 2012 (with the upgraded camera and at small zenith angles). The white circle indicates the size of VERITAS point spread function. The cross indicates the measured position of the TeV gamma-ray source. (right) Comparison of centroid positions from \textit{Fermi} (yellow, \cite{Yuan:2013cas}), VERITAS (green, \cite{VERITAS:2010cas}) and MAGIC (red, \cite{MAGIC:2003cas}) with the new VERITAS (white) centroid position.
}
\label{skymap}
\end{figure}

The VERITAS skymap of excess VHE gamma-ray events from the direction of Cas A is shown in Figure \ref{skymap} (left). This skymap is generated using the reflected background model \cite{Berge:2007} and smoothed with a circular window of radius 0.09\si{\degree}. Note that  the skymap shown here results from the analysis of 2012-2013 data (taken at small zenith angles) and corresponds to a statistical significance of 11 $\sigma$. The position of the gamma-ray source is determined by fitting a two-dimensional Gaussian function to a non-smoothed skymap. The blue cross on the skymap with RA=$23^{h}23^{m}20.4^{s} \pm 0\si\degree.006_{stat} \pm 0\si\degree.014_{sys}$ and Dec= $58.817 \pm 0\si\degree.006_{stat} \pm 0\si\degree.014_{sys}$ shows the centroid of the gamma-ray source. From this analysis, it is clear that position of the gamma-ray source is mainly limited by the systematic error in the pointing of the telescopes, even for this reduced dataset. In Figure \ref{skymap} (right), we shows the comparison of the centroid locations from the \textit{Fermi} (Yuan et al 2013 \cite{Yuan:2013cas}), VERITAS (Acciari et al 2010 \cite{VERITAS:2010cas}), MAGIC (Albert et al 2007 \cite{MAGIC:2003cas}) and VERITAS (new work). The new VERITAS centroid for the TeV emission is consistent with the center of the remnant, with the other TeV measurements, and with the Fermi-LAT position. The radius of GeV and TeV regions corresponds to 1 $\sigma$ statistical and systematic errors added in quadrature.

\begin{figure}[ht]
\begin{center}
\includegraphics [scale=0.5]{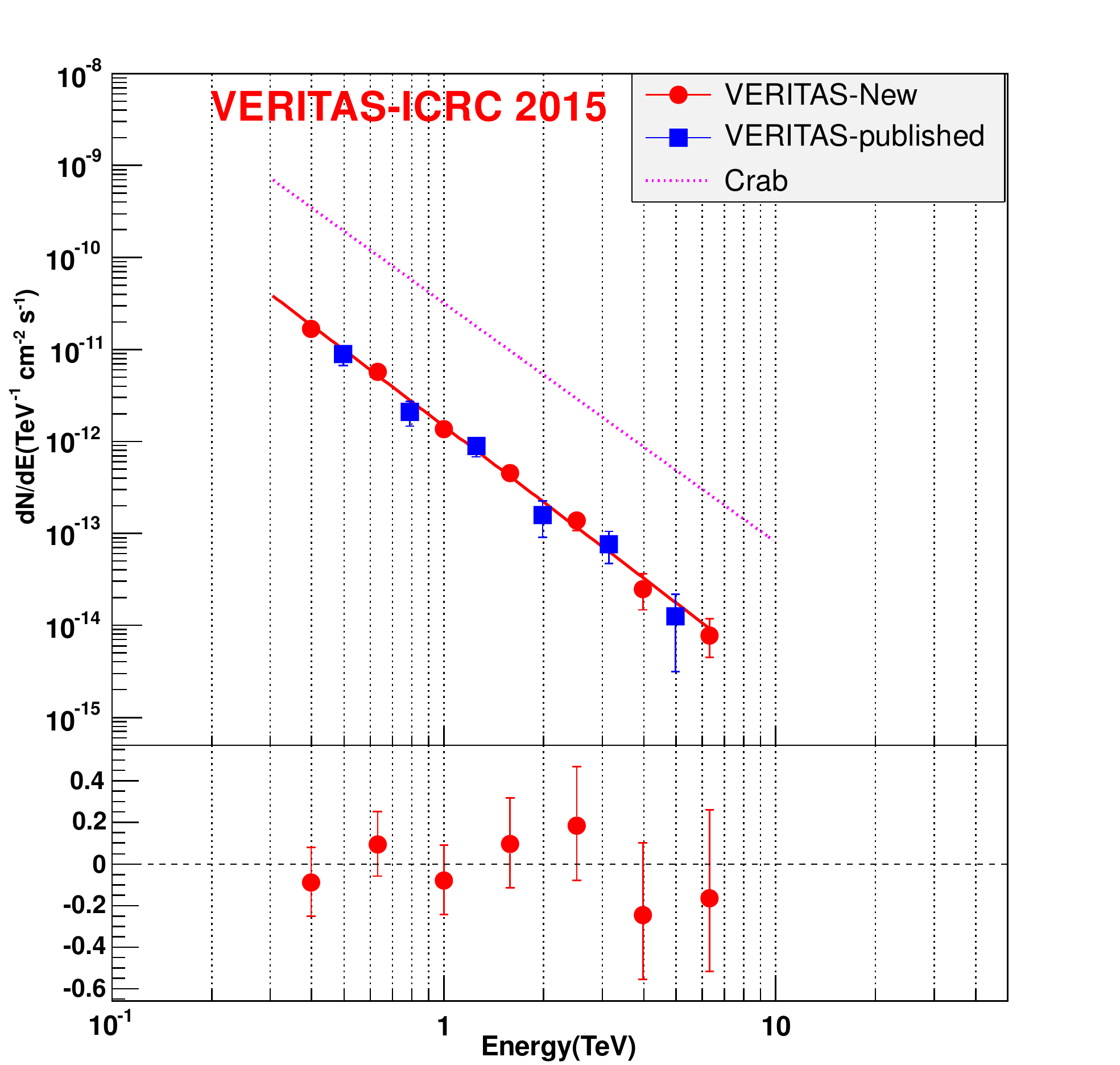}
\end{center}

\caption{\protect \small Differential energy spectrum from Cas A (this work) along with published VERITAS points (blue, Acciari et al 2010 \cite{VERITAS:2010cas}).
}

\label{energyspectrum}
\vspace{4mm}
\end{figure}

The differential energy spectrum for the whole data set  measured by VERITAS is also shown in Figure \ref{energyspectrum}, along with the published data points from VERITAS. The spectral points are fitted with a power-law in the energy range from 300 GeV to 7 TeV:
\begin{equation*}
\frac{dN} {dE} = (1.45 \pm 0.11) \times 10^{-12} ({E} / {1 TeV})^{-2.75 \pm 0.10_{stat} \pm 0.20_{sys}} cm^{-2} s^{-1} TeV^{-1}
\end{equation*}

A power-law fit to the data points gives a $\chi^{2}$ of 2.22 for 5 degrees of freedom, resulting in a good fit probability of 81\%.

\vspace{4mm}

\section{Summary and Discussions}
\begin{figure}[ht]
\begin{center}
\includegraphics [scale=0.7]{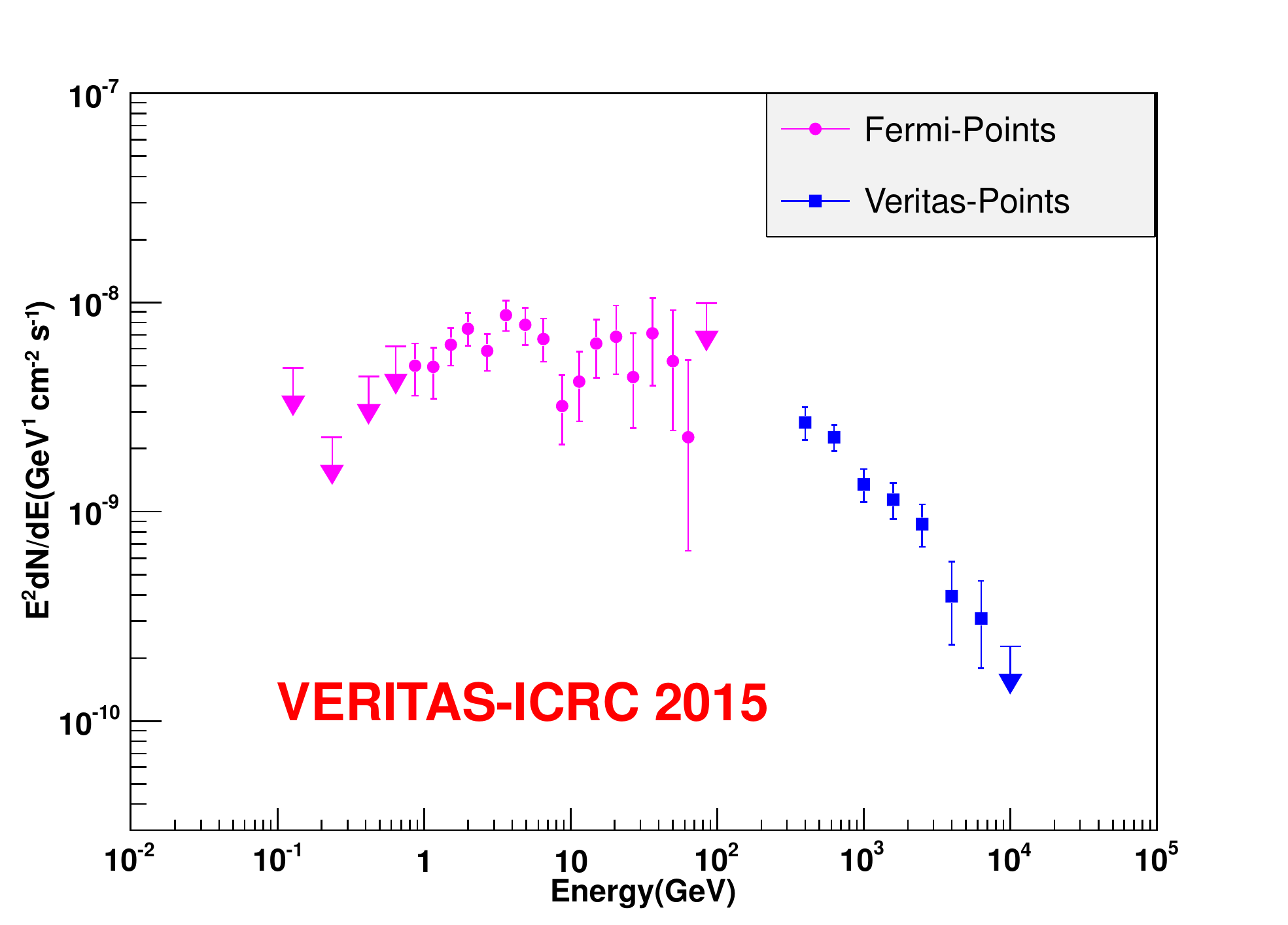}
\end{center}

\caption{\protect \small Combined differential flux points using \textit{Fermi} (Yuan et al 2013 \cite{Yuan:2013cas}) and VERITAS data (this work).
}
\label{fluxpoints}

\vspace{4mm}
\end{figure}

Cas A has been observed by VERITAS for $\sim$ 65 hours over a period of 6 years. One third of the total data has already been published \cite{VERITAS:2010cas}. By adding the new data to the already published data, we refine the spectral measurements (from our published spectrum) both at lower and higher energy. In addition to this, the improved signal-to-noise ratio helps to reduce the statistical errors on the spectral index by more than 50$\%$. For the centroid of the gamma-ray emission, we are now limited by the systematics in the pointing of our telescopes (50 arcseconds). 
Figure \ref{fluxpoints} shows the complete gamma-ray spectral energy distribution using the new flux points from \textit{VERITAS} and those from \textit{Fermi} \cite{Yuan:2013cas}. A discussion of these results will be presented at the conference.

\section{Acknowledgements}
This research is supported by grants from the U.S. Department of Energy Office of Science, the U.S. National Science Foundation and the Smithsonian Institution, and by NSERC in Canada. We acknowledge the excellent work of the technical support staff at the Fred Lawrence Whipple Observatory and at the collaborating institutions in the construction and operation of the instrument.
The VERITAS Collaboration is grateful to Trevor Weekes for his seminal contributions and leadership in the field of VHE gamma-ray astrophysics, which made this study possible.


\end{document}